# Visualization Tool: Exploring COVID-19 Data

*Dong Hyun Jeon, Jong Kwan Lee, Prabal Dhaubhadel, and Aaron Kuhlman*

**Dept. of Computer Science, Bowling Green State University, Bowling Green, Ohio 43403, U.S.A.**
*E-mail: { djeon | leej | pdhaubh | kuhlmaa }@bgsu.edu*

**Abstract**: The ability to effectively visualize data is crucial in the contemporary world where information is often voluminous and complex. Visualizations, such as charts, graphs, and maps, provide an intuitive and easily understandable means to interpret, analyze, and communicate patterns, trends, and insights hidden within large datasets. These graphical representations can help researchers, policymakers, and the public to better comprehend and respond to a multitude of issues. In this study, we explore a visualization tool to interpret and understand various data of COVID-19 pandemic.

While others have shown COVID-19 visualization methods/tools (e.g., [1]), our tool provides a mean to analyze COVID-19 data in a more comprehensive way. We have used the public data from NY Times [17] and CDC [18], and various COVID-19 data (e.g., core places, patterns, foot traffic) from Safegraph [19]. Figure 1 shows the basic view of our visualization view. In addition to providing visualizations of these data, our visualization also considered the Surprising Map [2]. The Surprising Map is a type of choropleth map that can avoid misleading of producing visual prominence to known base rates or to artifacts of sample size and normalization in visualizing the density of events in spatial data. It is based on Bayesian surprise—it creates a space of equi-plausible models and uses Bayesian updating to re-estimate their plausibility based on individual events.

## I. Introduction

As we convene in the midst of the COVID-19 pandemic, the critical role of precise data visualization is brought into sharp focus. This unprecedented global crisis has been characterized not only by its health implications but also by the sheer volume of data it has generated. Daily, a deluge of information is produced and disseminated, posing significant challenges in interpretation for policymakers, health professionals, and the broader public. Traditional data visualization methods, particularly thematic maps, are under scrutiny for their limitations, often leading to misinterpretations [3]–[6]. These conventional techniques struggle to encapsulate the complex and ever-evolving nature of pandemic data, thereby creating gaps in our understanding and response strategies.

Our work shifts focus to the development and utilization of advanced visualization tools. These tools are designed to not only accurately represent the intricate patterns of COVID-19 data but also to aid in informed decision-making. They are crucial in helping health professionals, policymakers, and the public navigate the multitude of challenges posed by the pandemic, ensuring that decisions are rooted in clear, comprehensive, and up-to-date data analysis. A cornerstone of our approach is the innovative use of Bayesian surprise to enhance the utility and accuracy of thematic maps. This method addresses the conventional pitfalls in data representation, such as base rate bias and renormalization, which can lead to misleading interpretations. In the context of the COVID-19 pandemic, where accuracy in data is paramount, our study applies the Bayesian surprise methodology to visualize COVID-19 data. We integrate diverse public data sources, including the NY Times, CDC, and Safegraph, to develop a comprehensive visualization tool. This tool surpasses the limitations of traditional choropleth maps by incorporating the 'Surprising Map' model. By emphasizing the unexpected elements in data, the Bayesian surprise model offers an accurate and insightful depiction of the pandemic's impact.

This paper is more than an exploration of a novel method; it represents a bridge between advanced statistical methods and practical data visualization techniques. It demonstrates the application of these methods to one of the most pressing issues of our era. Our aim is to enhance the clarity and accuracy of COVID-19 data visualization, thereby contributing to more informed decision-making and a deeper understanding of the pandemic's true impact.

## II. Related work

*A. Enhanced data analysis with visualization tools*

In the rapidly evolving field of data analysis, visualization tools have proven essential, especially in the context of the COVID-19 pandemic [3], [6], [7]. Research has highlighted the significant benefits of these tools in managing and interpreting large-scale pandemic data. Akhtar et al. [5] utilized Tableau for effective COVID-19 data visualization and analysis, demonstrating its capabilities in data blending, real-time reporting, and collaboration. Their methodology involved integrating Tableau with diverse data sources and creating interactive dashboards to enhance understanding and presentation of pandemic trends. Also, Dong et al. [8] introduced the COVID-19 Watcher, an application that visualizes data from every U.S. County and 188 metropolitan areas. This tool features rankings of severely affected areas and generates plots showing changes in testing, cases, and deaths, filling a critical gap in local data reporting, and providing essential real-time updates.

*B. The impact of various dashboard visualizations*

The design and functionality of dashboard visualizations significantly influence public perception, particularly during health emergencies like the COVID-19 pandemic [9], [10]. Studies have examined how different dashboard designs impact public health communication and risk perception. Naleef et al. [11] provided an in-depth analysis of COVID-19 dashboards used by state governments in the U.S., focusing on their design, functionality, and content alignment with public health goals. Padilla et al. [12] explored how various COVID-19 data visualizations affected public risk perception, conducting experiments with over 2,500 participants. They found that cumulative scale visualizations generally led to higher risk perceptions, whereas weekly incident scales resulted in more variable perceptions. Additionally, uncertainty forecast visualizations, particularly those incorporating multiple models, significantly increased risk estimates.

*C. Bayesian surprise in thematic map accuracy*

Accurate data representation in thematic maps is crucial across various fields, such as epidemiology and geography [13]–[16]. Correl et al. [2] delved into the application of Bayesian statistical methods to enhance thematic map accuracy. Their methodology centered on Bayesian surprise, which calculates the divergence between prior beliefs and observed data. This approach addresses biases commonly found in traditional thematic maps, like base rate bias and sampling error. It reinterprets how thematic maps represent data, leading to more accurate and insightful depictions of complex spatial distributions. The authors demonstrated the method's effectiveness in revealing underlying spatial patterns, using both synthetic and real-world datasets, and showcased its potential to uncover spatial anomalies that standard thematic maps might obscure.

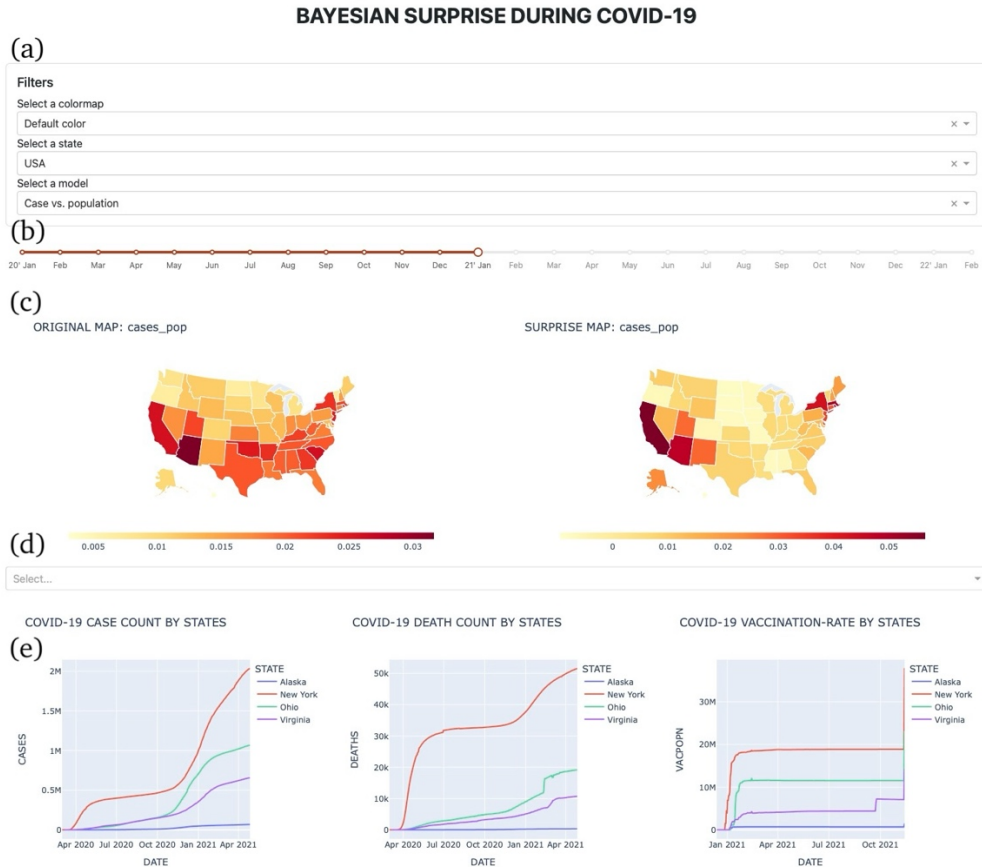

*Figure 1: Overview of the COVID-19 Visualization Tool. (a) Map color customization filters and options to select specific country regions by state or by larger areas like West, Midwest, South, and East. Feature to choose different data models for comparison. (b) Date range selector from January 2020 to February 2022 for data analysis. (c) Choropleth map to compare original and Bayesian surprise data. (d) Dropdown menu for state selection in analyzing COVID-19 statistics. (e) Line charts displaying case counts, death counts, and vaccination rates for the selected states.*

### III. Methodology

In our study, we employ the Bayesian surprise methodology to enhance the visualization of COVID-19 data. This approach, rooted in Bayesian statistics, offers a novel perspective in interpreting complex datasets, particularly in contexts where traditional methods might lead to misrepresentation or bias.

As illustrated in Fig 1, the first step involves the meticulous collection and preparation of COVID-19 data. We gather information from CDC and the NY Times, focusing on key metrics like infection rates, mortality rates, and demographic distributions. This data is then meticulously cleaned and standardized to ensure consistency and accuracy in our analyses. Once the data is prepared, we implement the Bayesian surprise model. This model calculates surprise as the Kullback-Leibler divergence between prior and posterior beliefs about the data, to measure how much new data changes our understanding compared to our previous beliefs. For this, we establish a Bayesian statistical model that represents our prior understanding of COVID-19's spread and impact. This model incorporates factors such as expected infection rates based on population density and demographic variables.

As new data comes in, the model updates its parameters, reflecting how the actual data differs from our prior expectations. These differences, quantified as 'surprises,' are crucial in identifying areas with unexpectedly high or low infection rates. By focusing on these surprises, we can highlight regions where the pandemic's impact defies conventional expectations, thereby providing deeper insights into the spread and effects of the virus. Moreover, our methodology extends to the visualization aspect. We employ choropleth maps enhanced with Bayesian surprise. These maps use color gradients not just to represent raw data, such as the number of cases, but also to visualize the degree of surprise. Areas with higher surprise values are rendered in more distinct colors, drawing attention to regions where the pandemic's behavior is notably different from expected patterns.

## IV. Visualization tool

W focused on evaluating the effectiveness of our advanced visualization techniques within the context of the COVID-19 pandemic. This involved a detailed exploration of various data sources, the intricate process of data preparation, and the establishment of a robust data pipeline, all of which were pivotal in facilitating our analyses.

### A. Data Sources

Our research utilized a diverse range of data sources, each offering unique and valuable perspectives on different facets of the pandemic. From SafeGraph, we accessed the Patterns dataset, which provided county-level monthly foot traffic information and the Core Places dataset for the number of Points of Interest (POIs) with their respective North American Industry Classification System (NAICS) codes. The raw data from SafeGraph was substantial, exceeding 150GB. Additionally, we incorporated data from The New York Times, renowned for its reliable and widely-used time-series data on COVID-19. This dataset, publicly available and compact in size (under 100 MBs in CSV format), added significant depth to our analysis. The Centers for Disease Control and Prevention (CDC) also played a crucial role, with their dataset providing vital vaccination-related information, approximately 150 MBs in size. Furthermore, to effectively synergize these varied datasets, we employed various lookup tables and crosswalks, such as FIPS to ZIP code mappings, and integrated census data for normalization.

### B. Data Preparation and Pipeline

The preparation and management of this data were central to our experimental process. Initially, all data, irrespective of its source, was acquired in CSV format. We then selected a MySQL server as our central repository, a choice motivated by its robustness and scalability. Handling the data required the adept use of Python's pandas library, particularly for addressing format issues like null values, thus ensuring data integrity. Following this, the data was loaded into the MySQL database for further processing. Here, we performed a range of manipulations and aggregations, with FIPS serving as the common column for all county-level data and state codes as the linking fields for state-level data.

### C. Visualization

The centerpiece of our data presentation is the application built on the Dash framework for Python. This framework is particularly adept for our purposes as it leverages PlotlyJS for graph generation, providing a high degree of interactivity and aesthetic control. To ensure the successful execution of this application, several dependencies are required to be met.

The design of the dashboards is oriented towards unifying all choropleth maps under a common timeframe selector. This allows for a cohesive and comprehensive view of the data across different metrics and regions. Key features like hover or tooltips enhance the user experience by providing detailed information on demand. The ability to scroll for zooming and the inclusion of a state-selector using checkboxes add layers of interactivity, enabling users to navigate and explore the data with ease. Another notable feature is chart-cropping, particularly useful for line charts. This functionality allows users to zoom into specific portions of a graph, facilitating a more detailed analysis of trends over time. The choice of colormaps and scaling methods was a critical aspect of our development process. These were carefully selected to ensure that they accurately represent the data and are intuitive for the user, enhancing the overall effectiveness of the visualization tool in communicating complex data patterns.

The results of these experiments were promising, indicating that our Bayesian surprise-based tool not only provided a more accurate representation of COVID-19 data but also enhanced user comprehension and engagement with the information presented. The tool effectively highlighted regions that required immediate attention or further investigation, thereby serving as an asset for decision-makers and the general public in understanding and responding to the pandemic.

## V. Conclusion

In conclusion, our study demonstrates the efficacy of the Bayesian surprise methodology in enhancing COVID-19 data visualization. Our tool not only provides a more accurate representation of the pandemic's impact but also improves user comprehension and engagement. The experiments reveal its ability to highlight critical yet overlooked data patterns, aiding in more informed decision-making. This approach signifies a major step forward in public health data analysis, offering a novel and effective means of understanding complex and dynamic datasets like those of COVID-19. Our findings underscore the potential of advanced statistical methods in transforming data visualization, especially in critical areas such as public health.

As a future work, rigorous validation of our model can be conducted. The Bayesian surprise maps with traditional thematic maps will be conducted to assess the added value and accuracy of our approach. This comparison involves not only statistical evaluation but also user studies, where we gather feedback from experts and laypersons on the interpretability and usefulness of our visualizations. By applying Bayesian surprise to COVID-19 data, we aim to provide a more accurate, insightful, and user-friendly tool for understanding the pandemic's complex spatial dynamics.

## Acknowledgment

We express our gratitude and recognition to Safegraph for being a principal source of our data, which was made accessible via Safegraph Academics.

## References

[1] K. Chakate, G. Giri, S. S. Gonge, A. Deshpande, Y. Pawade, and R. Joshi, "CoviCare: Tracking covid-19 using PowerBI," in *2022 8th international conference on signal processing and communication (ICSC)*, IEEE, 2022, pp. 74–78.
[2] M. Correll and J. Heer, "Surprise! Bayesian Weighting for De-Biasing Thematic Maps," *IEEE Trans. Vis. Comput. Graph.*, vol. 23, no. 1, pp. 651–660, Jan. 2017, doi: 10.1109/TVCG.2016.2598618.
[3] F. Clement, A. Kaur, M. Sedghi, D. Krishnaswamy, and K. Punithakumar, "Interactive data driven visualization for COVID-19 with trends, analytics and forecasting," in *2020 24th international conference information visualisation (IV)*, IEEE, 2020, pp. 593–598.
[4] E. Bowe, E. Simmons, and S. Mattern, "Learning from lines: Critical COVID data visualizations and the quarantine quotidian," *Big Data Soc.*, vol. 7, no. 2, p. 2053951720939236, 2020.

[5] N. Akhtar, N. Tabassum, A. Perwej, and Y. Perwej, "Data analytics and visualization using Tableau utilitarian for COVID-19 (Coronavirus)," *Glob. J. Eng. Technol. Adv.*, 2020.
[6] M.-F. Pang *et al.*, "Spatiotemporal visualization for the global COVID-19 surveillance by balloon chart," *Infect. Dis. Poverty*, vol. 10, pp. 1–8, 2021.
[7] Y. Zhang, Y. Sun, J. D. Gaggiano, N. Kumar, C. Andris, and A. G. Parker, "Visualization design practices in a crisis: behind the scenes with COVID-19 dashboard creators," *IEEE Trans. Vis. Comput. Graph.*, vol. 29, no. 1, pp. 1037–1047, 2022.
[8] E. Dong, H. Du, and L. Gardner, "An interactive web-based dashboard to track COVID-19 in real time," *Lancet Infect. Dis.*, vol. 20, no. 5, pp. 533–534, 2020.
[9] C. K. Leung, T. N. Kaufmann, Y. Wen, C. Zhao, and H. Zheng, "Revealing COVID-19 data by data mining and visualization," in *Advances in intelligent networking and collaborative systems: The 13th international conference on intelligent networking and collaborative systems (INCoS-2021) 13*, Springer, 2022, pp. 70–83.
[10] M. Shaito and R. Elmasri, "Map visualization using spatial and spatio-temporal data: Application to covid-19 data," in *The 14th PErvasive technologies related to assistive environments conference*, 2021, pp. 284–291.
[11] N. Fareed, C. M. Swoboda, S. Chen, E. Potter, D. T. Wu, and C. J. Sieck, "US COVID-19 state government public dashboards: an expert review," *Appl. Clin. Inform.*, vol. 12, no. 02, pp. 208–221, 2021.
[12] L. Padilla, H. Hosseinpour, R. Fygenson, J. Howell, R. Chunara, and E. Bertini, "Impact of COVID-19 forecast visualizations on pandemic risk perceptions," *Sci. Rep.*, vol. 12, no. 1, p. 2014, 2022.
[13] J. L. Comba, "Data visualization for the understanding of COVID-19," *Comput. Sci. Eng.*, vol. 22, no. 6, pp. 81–86, 2020.
[14] J. K. Teh *et al.*, "Multivariate visualization of the global COVID-19 pandemic: A comparison of 161 countries," *PLoS One*, vol. 16, no. 5, p. e0252273, 2021.
[15] A. H. M. Hassan, A. A. M. Qasem, W. F. M. Abdalla, and O. H. Elhassan, "Visualization & prediction of COVID-19 future outbreak by using machine learning," *Int. J. Inf. Technol. Comput. Sci*, vol. 13, no. 3, pp. 16–32, 2021.
[16] C. K. Leung, Y. Chen, C. S. Hoi, S. Shang, Y. Wen, and A. Cuzzocrea, "Big data visualization and visual analytics of COVID-19 data," in *2020 24th international conference information visualisation (iv)*, IEEE, 2020, pp. 415–420.
[17] NY Times COVID-19 Data, https://github.com/nytimes/covid-19-data (Lastly-accessed 2022).
[18] CDC Vaccination Data, https://cdc-vaccination-history.datasette.io/ (Lastly-accessed 2022).
[19] Safegraph Data: Core Places, Patterns, and Geometry, https://www.safegraph.com (Lastly-access 2022).